\documentstyle[11pt]{article}
\newcommand{\comment}[1]{}
\newcommand{\s}{_{st}}

\begin{document}

\title{The Parallel Complexity of Growth Models}

\author{Jonathan Machta
\thanks{This research was partially funded by the National
Science Foundation Grant DMR-9014366 and 9311580.}\\
Department of Physics and Astronomy\\
University of Massachusetts\\
Amherst, Massachusetts 01003\\
e-mail address: machta@phast.umass.edu
\and
Raymond Greenlaw
\thanks{This research was partially funded by the National
Science Foundation Grant CCR-9209184.} \\
Department of Computer Science \\
University of New Hampshire \\
Durham, New Hampshire 03824\\
e-mail address: greenlaw@cs.unh.edu\\[.5in]
}

\maketitle

\begin{abstract}
This paper investigates the parallel complexity of several
non-equilibrium growth models. {\em Invasion percolation}, {\em Eden
growth}, {\em ballistic deposition\/} and {\em solid-on-solid
growth\/} are all seemingly highly sequential processes that yield
self-similar or self-affine random clusters. Nonetheless, we present
fast parallel randomized algorithms for generating these clusters.
The running times of the algorithms scale as $O(\log^2 N)$, where $N$ is the
system size, and the number of processors required scale as a polynomial in
$N$\@. The algorithms are based on fast parallel procedures for finding minimum
weight paths; they illuminate the close connection between growth
models and self-avoiding paths in random environments.  In addition to
their potential practical value, our algorithms serve to classify
these growth models as less complex than other growth models, such as
{\em diffusion-limited aggregation\/}, for which fast parallel
algorithms probably do not exist.
\end{abstract}


\newpage

\section{Introduction}
\label{se-intro}

Non-equilibrium growth phenomena are encountered in a diverse array of
physical settings~\cite{KrSp,Vi,FaVi}.  In many cases, non-equilibrium
growth leads spontaneously to complex self-affine or self-similar
patterns. A variety of models have been developed to gain an
understanding of how complex patterns develop on large length scales
due to the operation of simple microscopic rules.  Although much is
known about these models, most of our insight has been derived from
numerical simulation.  Theoretical analysis parallels the theory of
equilibrium critical phenomena but is less well developed.  For
example, it is believed universality classes exist~\cite{KrMeHa} that
include diverse models all having the same scaling exponents, but
theoretical arguments are generally less compelling than in the
equilibrium case.

In the present paper we examine several well-known growth models from
the perspective of parallel computational complexity.  Specifically we
study {\em invasion percolation}, {\em Eden growth}, {\em ballistic
deposition\/} and {\em solid-on-solid growth}, and show that fast
parallel algorithms exist for all of these models.  All of the
algorithms make use of random numbers.  Here `fast' means that the
running time scales polylogarithmically ($O(\log^{k_1}N)$ for some
constant $k_1$ and abbreviated {\em polylog}) in the size of the
system and that the number of processors scales polynomially
($O(N^{k_2})$ for some constant $k_2$) in the system size.  It is
interesting to compare our results with a previous one~\cite{Mac93a}
that proves it is very unlikely there is a fast parallel algorithm for
{\em diffusion-limited aggregation\/} (DLA).  The models studied here
are much more amenable to parallelization than DLA.

There are several reasons for investigating growth models from the
point of view of parallel computation.  First, with the rapidly
increasing availability of massively parallel computers it is
important to develop approaches for simulating various physical
systems in parallel. Although the algorithms presented here are for an
idealized parallel model, they may serve as a starting point for the
design of practical algorithms for large scale parallel machines.  The
approach described here minimizes the running time by using a
polynomial number of processors with each processor doing very little
computational work but having unlimited interconnections.  If our
algorithms were implemented directly on a parallel machine, the
processors we require would be treated as virtual processors.  Each
physical processor would have to perform the work of many virtual
processors (as is typically done now with algorithms that require more
than the available number of physical processors).  For different
parallel architectures, the running times would vary depending on the
underlying communication network of the machine.  It should be
possible to achieve significant speed up using the approaches
described here; these techniques demonstrate the unexpected inherent
parallelism in the problems we consider.

On a more fundamental level, the existence, structure and complexity
of the parallel algorithms provides a new perspective on the models
themselves.  One question that can be addressed is whether a growth
process is intrinsically history dependent.  All of the growth models
discussed here are apparently history dependent in the sense that the
growth at a given time depends on the prior history of the system.
For example, consider the growth rules for Eden clusters.  At each
time step a new particle is added randomly to the perimeter of the
cluster. This particle modifies the cluster and its perimeter.  The
model is defined as a sequential procedure that requires
$K$ steps to create a cluster with $K$ particles. It is not at
all obvious that one can generate randomly chosen Eden clusters in
polylog time even with many processors running in parallel.  The
existence of fast parallel algorithms for the growth models considered
here shows that their apparent history dependence can be overcome.

Invasion percolation yields self-similar clusters and is in the same
universality class~\cite{ChChNe} as static percolation~\cite{StAh} at
the percolation threshold.  Eden growth, ballistic deposition and
solid-on-solid growth yield self-affine surfaces and are believed to
be in the universality class described by the continuum Kardar, Parisi
and Zhang (KPZ) equation~\cite{KPZ}.  The directed polymer in a random
environment~\cite{KaZh} is also in this universality class.  Both
static percolation and the directed polymer are equilibrium models
with fast parallel algorithms; their algorithms are specified in
Section~\ref{su-mwpath}.  Together these results suggest that the
non-equilibrium growth models may also be amenable to fast parallel
simulation.

The key step in constructing parallel algorithms for the
non-equilibrium growth models is mapping them onto minimum weight path
problems for which fast parallel algorithms already exist.  Eden
growth and the solid-on-solid model have been
shown~\cite{RoHaHi,TaKeWo} to be equivalent to a {\em waiting time\/}
growth model that can then be mapped onto a minimum weight path
problem. We show that invasion percolation is also equivalent to a
waiting time growth model with a different distribution of waiting
times.  We also give a new connection between the growth models in the
KPZ universality class and minimum weight paths that leads to a
distinct group of algorithms using a {\em random list\/} approach.
The parallel implementations of the random list and waiting time
approaches result in different running times, processor requirements
and use of randomness.

The remainder of this paper is organized as follows: in
Section~\ref{se-parallel} we present some background material on
parallel computation and describe the {\em parallel random access
machine\/} (P-RAM) model of computation that our algorithms
are defined on, in Section~\ref{se-growth} we describe the growth
models and relate them to minimum weight path problems, in
Section~\ref{se-parallelgrowth} fast parallel algorithms are developed
for the growth models using the random list and random weight
approaches, and in Section~\ref{se-discussion} we discuss the results.

\section{Parallel computation}
\label{se-parallel}

In this section we describe the P-RAM model of parallel computation,
present some background on parallel complexity theory, outline fast
parallel algorithms for the minimum weight path problem and for
generating random permutations, and define a probabilistic variant of
the P-RAM.

\subsection{The P-RAM model of parallel computation}
\label{su-pram}

The theory of parallel computation can be based upon several different
but nearly equivalent abstract computational models.  The model used
here and in the bulk of the computer science literature is the
P-RAM\@.  We present an informal description of the model below.  The
P-RAM consists of a number of processors each with local memory and
access to a common global random access memory.  Each memory contains
an unlimited number of cells.  The processors use the same program and
are distinguished by integer labels.  Input to the machine is placed
in designated, consecutive global memory locations as is the output.
The P-RAM is in the class of single-instruction multiple-data-stream
(SIMD) models.  The processors run synchronously and in each time step
a single {\em random access machine\/} (RAM) instruction~\cite{AhHoUl}
is executed by some of the processors.  Examples of typical
instructions are `write the contents of the accumulator to memory
location $a$' and `add the contents of the accumulator and the
contents of register $a$, placing the result in the accumulator.'
Although many processors may read the same memory location at the same
time, difficulties arise if multiple processors attempt to write to
the same location.  Here we adopt the concurrent read, exclusive write
or CREW P-RAM model in which only one processor is allowed to write to
a given memory location at a time.  In this model if two processors
attempt to write to the same memory cell, the program fails.  Write
arbitration schemes are discussed in~\cite{Fi}.

In the P-RAM model any processor can access any global memory location
in one time step; the model allows unlimited parallelism.  The P-RAM
is an idealized model which allows easy expression of algorithms and
provides a convenient framework for addressing new problems.  The P-RAM
is also useful for proving lower bounds.  There are general simulations
showing that the P-RAM can be simulated on more realistic parallel
computers.  These simulations usually have a slow down of a logarithmic
factor and require the same amount of hardware as the corresponding P-RAM
computations, see~\cite{KaRa} for additional details and references.

\subsection{Parallel complexity theory}
\label{su-pct}

Parallel complexity theory seeks to determine the time and processor
requirements for solving computational problems in parallel, and to
develop practical parallel algorithms.  A first step towards
understanding how well a problem parallelizes is to determine how the
time and processor requirements (on an idealized model) scale with the
size of the problem, where `size' refers to the number of bits needed
to specify the problem in some parsimonious form.  We illustrate some
basic definitions in complexity theory via the problem of finding the
minimum among $N$ natural numbers, $k_1,\ldots,k_N$, where the numbers
in the list are bounded by $B$\@.  The input to the problem can be
specified by $N\log B$ bits where each group of $\log B$ bits is
interpreted as one number.  It is clear that a sequential computer,
such as a RAM or a more familiar desktop computer with sufficient
memory, could solve this problem in $O(N)$ steps.  A single for loop
that maintains the minimum of the numbers scanned thus far produces
the overall minimum after considering each number in the list.

Finding the minimum of $N$ numbers is thus in the complexity class,
{\bf P}, consisting of all problems that can be solved on a sequential
computer in a time bounded by a polynomial in the problem size.  How
would one find the minimum on a P-RAM?\footnote{In the RAM and P-RAM
models, registers and memory locations may hold unbounded integers.
Instructions may operate on unbounded integers.  The {\em log cost
model\/} is used to obtain a realistic cost of how many registers are
actually needed to store a large value and to gauge the time of an
instruction involving large values.  See~\cite{AhHoUl} for further
details.} For simplicity, suppose the length of the input list is a
power of two, i.e. $N=2^p$ for some natural number $p$.  Assume the
input is stored in the $N$ memory locations, $M[i], i=1,2,\ldots,N$\@.
A parallel algorithm to find the minimum is given below.

\begin{tabbing}
{\bf Parallel Minimum Algorithm}\\
\hspace*{.25in} {\bf beg}\={\bf in}\\
\>{\bf for} \= $l \leftarrow 1$ {\bf to} $\log N$ {\bf do}\\
\>\> {\bf for} \= {\bf all} $i, 1 \leq i \leq N/2^{l-1}$,
{\bf in parallel do} \\
\>\>\>{\bf if} $M[2i-1] \leq M[2i]$ {\bf then}
$M[i] \leftarrow M[2i-1]$ {\bf else}
$M[i] \leftarrow M[2i]$;\\
\hspace*{.25in} {\bf end}.
\end{tabbing}

\noindent
At each time step, adjacent odd and even positions of $M$ are compared
and the minima are written to the first half of the remaining
$M[i]$'s.  Clearly, the running time of this algorithm is $O(\log N)$
and $N$ processors are needed.  We refer to this as a {\em fast\/}
parallel algorithm since it solves the problem in polylog time with
polynomially many processors.  The algorithm sketched above is not an
{\em optimal\/} parallel algorithm since the product of the time and
number of processors exceeds, by a log factor, the time requirement
for the optimal sequential algorithm.  In the present case it is easy
to find an optimal parallel algorithm that runs in $O(\log N)$ time
but with only $N/\log N$ processors.  A detailed presentation of fast
parallel algorithms for a variety of tasks is given in~\cite{GiRy}.

The class of problems that can be solved on a P-RAM in polylog time
with polynomially many processors is called {\bf NC}\@.  Clearly,
${\bf NC} \subseteq{\bf P}$ and it is strongly believed that the
inclusion is strict, implying there are problems in {\bf P} that
cannot be solved in parallel in polylog time using a polynomial number
of processors. A class of `{\bf P}-complete' problems~\cite{GrHoRu}
can be identified that are representative of the hardest problems to
solve in parallel (in polynomial time using a polynomial number of
processors).  The definition of {\bf P}-completeness and the unproved
conjecture that {\bf NC}$\neq${\bf P} is analogous to the
better known theory of {\bf NP}-completeness~\cite{GaJo} and the famous
unproved conjecture that {\bf P}$ \neq ${\bf NP}\@.  Given the
assumption {\bf NC}$\neq${\bf P}, {\bf P}-complete problems cannot be
solved in polylog time using a polynomial number of processors;
thus they have an intrinsic history dependence.  Previous
work~\cite{Mac93a} showed that a natural problem yielding DLA clusters
is {\bf P}-complete, whereas the growth models considered here are in
the complexity class random {\bf NC}, the randomized version of {\bf
NC} that is denoted {\bf RNC}\@.  This suggests that DLA is
intrinsically history dependent and, in a fundamental sense, more
complex than the growth models discussed here.

The distinction between polynomial parallel time and polylog parallel
time is a robust one; it is independent of the model of computation.
Indeed the distinction between complexity classes is perhaps most
perspicuous in a very different computation model consisting of
`circuit families.'  A {\em Boolean circuit\/} is comprised of {\sc
AND}, {\sc OR} and {\sc NOT} gates; connecting wires, inputs and
outputs.  {\sc AND} and {\sc OR} gates have fan-in two and fan-out two
while {\sc NOT} gates have fan-in one and fan-out two.  The gates are
connected without loops and the circuit computes outputs from its
inputs in the obvious time ordered way.  The {\em depth\/} of a
circuit is the length of the longest path from an input to an output.
Depth corresponds roughly to the time of a parallel computation.  The
{\em size\/} of the circuit is the number of gates appearing in the
circuit.  Size corresponds roughly to the computational {\em work}, or
the number of steps required to solve the problem by a sequential
computer.  To solve a class of problems of varying size we need a
family of circuits, one for each input size. A circuit family is {\em
uniform\/} if the members of the family are structurally
related.\footnote{A circuit family is {\em logspace uniform\/} if the
design of the $n^{\rm th}$ circuit in the family can be constructed by
some Turing machine on input $1^n$ using $O(\log C(n))$ workspace, where
$C(n)$ is the size of the circuit.  In the following `uniform' means
`logspace uniform.'} The reader is referred to~\cite{GrHoRu} for a
discussion of uniformity.

Computational complexity classes can be defined for circuit families
in terms of how the size and depth scales with the problem size.  Any
problem that can be solved by a P-RAM in polylog time with
polynomially many processors can be solved by a uniform circuit family
with polylog depth (the bounded fan-in requirement implies the circuit
has polynomial size).  Thus the complexity class {\bf NC} consists of
problems solvable by uniform circuit families with polylog depth.  One
can define subclasses within {\bf NC}; the class {\bf NC}$^k$, $k >
1$, consists of those problems that can be solved by uniform circuit
families with bounded fan-in having depth $O(\log^k N)$.  Note, there
are additional technical considerations for the case $k$ equals 1.  A
problem that is solvable by a CREW P-RAM in $O(\log^k N)$ with a
polynomial number of processors can be solved by a uniform circuit
family with depth $O(\log^{k+1} N)$. The randomized version of
{\bf NC}$^k$ is denoted {\bf RNC}$^k$.  We describe how randomness is
introduced into the models in Section~\ref{su-ppram}.

\subsection{The minimum weight path parallel algorithm}
\label{su-mwpath}

At the heart of the fast parallel algorithms for growth models is a
standard {\bf NC}$^2$ algorithm for finding minimum weight paths
(MWPs) between each pair of vertices in an
undirected\footnote{Throughout the paper for two sites labeled $i$ and
$j$, an undirected edge or bond between $i$ and $j$ is denoted by
$\{i,j\}$ and a directed edge or bond from $i$ to $j$ is denoted by
$(i,j)$.} graph~\cite{GiRy}.  The MWP problem is defined below.

\begin{tabbing}
{\bf Output:} \= Minimum weight simple paths connecting every pair of \kill
{\bf Input:} \> An undirected graph, $G=(V,E)$,
where $V$ is a set of sites and $E$ is a
set of bonds \\ \> connecting pairs of sites.
Non-negative
weights, $w(i,j)$,
assigned to each bond, $\{i,j\} \in E$. \\
{\bf Output:} A matrix containing weights of the minimum weight {\em
simple paths\/} between every \\ \> pair of sites in $V$\@.  A simple
path is a connected sequence of edges without cycles. \\ \> The
weight of a path is the sum of the weights of its edges.
\end{tabbing}

Let $N = |V|$.  A parallel algorithm to solve the MWP problem is given
below.  It takes as input the matrix of weights, $w(i,j)$, $1 \leq i,j
\leq N$, with $w(i,i)=0$\@.  The output matrix $W(i,j)$ holds the weights
of the paths.

\begin{tabbing}
{\bf Parallel Minimum Weight Path Algorithm}\\
\hspace*{.25in} {\bf beg}\={\bf in}\\
\> {\bf for} \= $i,j$, \=$1 \leq i,j \leq N$, {\bf in parallel do}\\
\>\>{\bf if} ($\{i,j\} \in E$ {\bf or} $i = j$)
{\bf then} $W(i,j) \leftarrow w(i,j)
$ {\bf else} $W(i,j) \leftarrow \infty$;\\
\>{\bf for} \= $l \leftarrow 1$ {\bf to} $\lceil \log N \rceil$ {\bf do}\\
\>\> {\bf for} \= {\bf all} $i,j,k$, $1 \leq i,j,k \leq N$,
{\bf in parallel do}\\
\>\>\> $W(i,j) \leftarrow \min_k[W(i,k)+W(k,j)]$;\\
\hspace*{.25in} {\bf end}.
\end{tabbing}

\noindent
Note in the last step the minimum is over all $k$ in the range $1,2,
\ldots, N$\@.  At the $l^{\rm th}$ step in the outer for loop,
$W(i,j)$ contains the weight of the MWP from $i$ to $j$ having length
$2^l$ or less.  Each step allows the path length to double.  Since
finding the minimum over $N$ numbers can be done in $\log N$ time and
since the iteration is repeated $\log N$ times, the parallel MWP
algorithm takes $O(\log^2 N)$ time.  It requires $N^3/ \log N$
processors on the CREW P-RAM~\cite[page 26]{GiRy}.  Note that this is
not an {\em efficient\/} algorithm since the best sequential algorithm
requires $O(N^2)$ steps (technically, $O(N \log N + |E|)$,
see~\cite[page 920]{KaRa}); thus it does polynomially less
computational work.  The resources required by the MWP algorithm
dominate those needed for fast simulations of the growth models as we
will see in Section~\ref{se-parallelgrowth}.

The parallel MWP algorithm can be used to find path weights for which
a quantity other than the {\em sum\/} of the weights is either
minimized or maximized. Let $\oplus$ stand for an associative binary
operation such as addition, multiplication or taking the maximum.  If
the weight of a path is defined by the $\oplus$ of the weights along
the path, the MWP problem can be solved by the parallel MWP algorithm
by simply replacing $+$ by $\oplus$.

The parallel MWP algorithm can be used for directed graphs by changing
the braces to parentheses in line three.  It is also easily adapted to
site weights.  Here the weight of the destination is included in the
path weight but the weight of the source is not included. Finding
minimum weight paths on an undirected graph with site weights, $w(i)$,
is accomplished by constructing a corresponding directed graph with
bonds $(i,j)$ and $(j,i)$ for each undirected bond $\{i,j\}$.  A bond
weight $w(i,j)=w(j)$ is assigned to each bond $(i,j)$ and the
algorithm is run for this directed bond weight problem.  The
corresponding MWP values for the undirected graph having site weights
are given by the $W_d(i,j)$'s, where $W_d(i,j)$ is the value computed
in the directed bond weight problem.  The algorithm can also be used
to find the {\em connected components\/} of a graph.  A connected
component (for an undirected graph) is a set of sites for which a path
exists between every pair of sites.  To find the connected component
containing site $k$, the weight function is chosen so that $w(i,j)=0$
for all edges $\{i,j\}$, $1 \leq i,j \leq N$\@.  If $W(k,j)=0$, $j$ is
in the component connected to $k$.

\subsubsection{Static percolation and the directed polymer}

Two equilibrium problems related to the growth models, static
percolation and the directed polymer in a random environment, can be
solved by the parallel MWP algorithm.  Static percolation clusters are
connected components and thus can be identified by the MWP algorithm
as described above.  At zero temperature the directed polymer is a MWP
with $w(i,j)$ representing the energy of the directed bond $(i,j)$.
At finite temperature the partition function is the weighted sum over
paths that can be computed by replacing

\begin{center}
$\min_k[W(i,k)+W(k,j)] \; \;$ by $\; \; \sum_k[W(i,k)\times
W(k,j)]$
\end{center}

\noindent
in the parallel MWP algorithm and letting $w(i,j)$ be the Boltzmann
weight for bond $(i,j)$.

\subsection{Probabilistic parallel computation}
\label{su-ppram}

The natural problems in computational statistical physics are {\em
sampling\/} problems.  The objective is to generate an unbiased random
sample from a distribution of system states or histories. This is
somewhat different from the {\em decision\/} problems usually
considered in complexity theory. Decision problems are problems having
``yes'' or ``no'' answers.  The algorithms we describe produce a list
of integers as output.

In order to generate random objects one must have a supply of random
or pseudo-random numbers.  We assume a supply of perfect random
numbers although in practice a good pseudo-random number generator
would be used.  The probabilistic model we adopt for studying sampling
problems is a variant of the P-RAM\@. The {\em probabilistic P-RAM\/}
is a P-RAM in which each processor is equipped with a register for
generating random numbers.  If a natural number, $M$, is supplied to
the register, a random number in the range $1,2, \ldots, M$ is
returned.  Let RANDOM($M$) be the result of a call to this
register. This is the same model as adopted in~\cite{Ha}.  Typically,
it is assumed that $M =O(N)$, where $N$ represents the input size of
the problem under consideration.\footnote{The register generating
random numbers returns random bits rather than random numbers in some
variants of the model.} For the probabilistic P-RAM the generation of
random numbers is a primitive operation rather than the output of a
pseudo-random number generator.

A simple example of a sampling problem is to produce random
permutations of $N$ objects.  Random permutations will be needed for
generating invasion percolation clusters in parallel, in
Section~\ref{su-pip}.  The standard sequential algorithm for
generating a random permutation is sketched below.

\begin{tabbing}
{\bf Sequential Random Permutation Generation Algorithm}\\
\hspace*{.25in} {\bf beg}\={\bf in}\\
\>{\bf for} \= $i \leftarrow 1$ {\bf to} $N$ {\bf do}\\
\>\> $\Pi[i] \leftarrow i$;\\
\>{\bf for} $i \leftarrow N$ {\bf down to} $2$ {\bf do}\\
\>\> $j \leftarrow$ RANDOM$(i)$;\\
\>\> swap$(\Pi[i], \Pi[j])$;\\
\hspace*{.25in} {\bf end}.
\end{tabbing}

One way to view this algorithm is as a composition of $N - 1$
permutations.  Suppose array $\Pi$ is initialized as above and let $j
\leftarrow$ RANDOM$(i)$.  Let $\pi_i$, $2 \leq i \leq N$, be the
permutation formed by swapping $i$ and $j$ in $\Pi$ while leaving all
other elements fixed.  The composition $\pi_2 \circ \cdots \circ
\pi_N$ is a random permutation.  It is now easy to see how to
parallelize the computation using a binary tree of height $O(\log N)$.
The only problem is how to compose two arbitrary permutations.
In~\cite{Ha} a fast way of composing permutations is given, and an
$O(\log N)$ time, $N$ processor algorithm on an EREW P-RAM (exclusive
read, exclusive write) results for producing a random permutation.
The procedure requires $O(N \log N)$ bits of random information.

The complexity of sampling problems can also be discussed in terms of
{\em probabilistic circuit families}. Probabilistic circuits are usual
Boolean circuits supplemented with random sources.  The sources have
zero inputs and $M$ outputs.  When a random source is used, one of the
$M$ outputs is chosen (independently) with probability $1/M$ and takes
the value TRUE while all of the other outputs assume the value
FALSE\@.\footnote{Other models of probabilistic circuits permit
sources of random bits only.} The {\em depth\/} and {\em size\/} of
probabilistic circuits are defined analogously as they were for
typical Boolean circuits.  In the language of circuit complexity, our
results show there are probabilistic circuit families of polylog depth
and polynomial size that are capable of generating clusters for each
of the growth models.  Although it would be impractical to build such
devices, it is interesting that special purpose logic circuits, which
can make random choices, could be built that generate, say Eden
clusters on graphs of size $N$\@.  Further, the parallel running time
of these devices is only polylog in $N$.

Technically, the complexity class {\bf NC}$^k$ refers to decision
problems rather than sampling problems.  If however, the outputs of
the random sources used by the probabilistic circuits are treated as
inputs to deterministic circuits, we can speak of the deterministic
part of our algorithms as solving problems in {\bf NC}$^k$.

\section{Growth models defined on general graphs}
\label{se-growth}

Growth models are conventionally defined on $d$-dimensional lattices,
however, it is convenient for our purposes to define these models
on more general structures, i.e. graphs.  The graph theoretic
viewpoint simplifies the presentation of the parallel algorithms and
allows us to make connections with algorithms for other graph theory
problems.  The reader is referred to~\cite{Ev} for an introduction to
graph theoretic language and notation.

For invasion percolation and the Eden model, growth proceeds on an
undirected graph, $G=(V,E)$.  The output of the models is a connected
set of sites called a {\em cluster\/} together with an ordering,
$c:{\bf K} \rightarrow V$, of when sites are added to the cluster.
Here ${\bf K}= \{0,1,\ldots, K\}$, where $K$ denotes the size of the
cluster.  In general, $K$ will denote the number of iterations of a
growth rule.  The function $c$ will specify an ordering.  For a fixed
graph each growth model defines a different probability distribution
on the cluster ordering.  The distributions are defined by rules for
how new sites are added to the cluster.  The new site is chosen from
the {\em perimeter\/} of the existing cluster.  A perimeter site is a
non-cluster site that is connected by some bond to an element of the
cluster.  Growth begins with an initial cluster $S$.

The ballistic deposition and solid-on-solid models describe
directional growth.  An undirected graph, $G=(V,E)$, defines a
`substrate' above which growth occurs. Growth occurs on a `space-time'
graph, $G\s=(V\s,E\s)$, derived from $G$\@.  The sites, $V\s = \{ [i,h]
\; \mid \; i \in V$ and $h=0,1, \ldots, H\}$, form columns and $h$ is
interpreted as the height above the substrate.  The directed
bonds for the space-time graph, $E\s$, are specified below in
Section~\ref{su-sos}.

Each of the four models described above has been extensively studied
and a wide range of variants have been introduced to model specific
physical situations or to improve numerical results.  Below we
describe the basic versions of the growth rules for each model.

\subsection{Invasion percolation and invasion percolation with trapping}
\label{su-ip}

Invasion percolation and invasion percolation with
trapping~\cite{WiWi,ChKo} are defined via deterministic growth rules
operating in a system with quenched randomness.  Each site, $j \in V$,
is independently assigned a random number, $x_j$, chosen from a
uniform distribution on $[0,1]$.  The cluster starts from an initial
set and grows by the addition of the perimeter site with the smallest
random number

\begin{tabbing}
\hspace*{.5in} CLUSTER $\leftarrow$ CLUSTER $\cup \; \{j\}$,
where $j$ is the perimeter site with the smallest value of $x_j$;
\end{tabbing}

\noindent
The growth rule is iterated $K$ times and $c(n)$ denotes the $n^{\rm
th}$ site added to the cluster.  Note, by convention $c(0) = S$.

It is known~\cite{ChChNe} that on large lattices, the invasion
percolation process has properties that are closely related to
ordinary percolation at the percolation threshold~\cite{StAh,Ke}.  The
clusters are statistically self-similar with a Hausdorf dimension less
than the underlying space dimension.

Invasion percolation was devised to model one fluid (the invading
fluid) displacing a second fluid (the defending fluid) in a porous
material.  The random numbers reflect the random barriers due to
surface tension obstructing the growth of the cluster of invading
fluid.  If the defending fluid is incompressible, it must have a flow
path to leave the system through a {\em sink\/} site, $s'$.
Therefore, regions of defending fluid that become disconnected from
the sink are not invaded.  This constraint is embodied in the growth
rule for invasion percolation with trapping

\begin{tabbing}
\hspace*{.5in} \= CLUSTER $\leftarrow$ CLUSTER
$\cup \; \{j\}$, where $j$ is the perimeter site with the smallest
value of $x_j$\\
\>connected to $s'$ by a simple path that is disjoint
from the cluster;
\end{tabbing}

\noindent
The addition of the trapping rule adds a non-local constraint to the
invasion percolation algorithm.  It was originally believed that
invasion percolation with trapping was in a different universality
class than invasion percolation and static percolation.  However, it
is now known that all three models are in the same universality
class~\cite{PoNeMe}.  We shall see that few modifications are required
to incorporate the trapping rule into the fast parallel algorithm for
invasion percolation.

\subsection{Eden growth}
\label{su-eg}

Eden growth was originally introduced to simulate tumor
growth~\cite{Eden,KrSp}. The growth rules are similar to invasion
percolation except that randomness is introduced dynamically

\begin{tabbing}
\hspace*{.5in} \= CLUSTER $\leftarrow$ CLUSTER
$\cup \; \{j\}$, where $j$ is chosen with equal
probability from the set of \\
\> perimeter sites of the cluster;
\end{tabbing}

\noindent
Unlike invasion percolation, large Eden clusters grown on lattices are
compact~\cite{Ri} and nearly spherical but exhibit statistically
self-affine surface roughness~\cite{PlRa,FrStSt}. It is believed that
the exponents characterizing this surface roughness are the same as
those of the next two models we describe and also of the stochastic
differential equation introduced by KPZ~\cite{KPZ}.

\subsection{Ballistic deposition}
\label{su-bd}

Ballistic deposition~\cite{Vold,MeRaSaBa,KrSp} was developed to model
sedimentation, colloidal aggregation and vapor deposition of thin
films.  Ballistic deposition simulates growth above a substrate that
is here taken as an undirected graph, $G = (V,E)$.  Each site of the
graph defines a `column.'  A function, $h(i,n)$, is interpreted as the
height of column $i$ at step $n$. Initially, the height of each site
is set to zero and at each step a site, $i$, is randomly chosen.  The
height of $i$ is incremented according to the following rule:

\begin{tabbing}
\hspace*{.5in}
$h(i,n+1)\leftarrow \max [
\{ h(j,n) \; | \; \{i,j\} \in E \} \cup \{ h(i,n)+1 \} ]$;
\end{tabbing}

\noindent
If the substrate is a two-dimensional lattice, ballistic deposition
simulates particles falling one at a time and sticking at the highest
level at which they meet the growing cluster.  In terms of the
space-time graph, $G\s$, described at the beginning of
Section~\ref{se-growth}, the initial cluster is the $h(i,0)=0$
`plane.'  At each step the cluster grows at the site where the height
is incremented, i.e. $c(n)= [i,h(i,n)]$ if $h(i,n)$ is greater than
$h(i,n-1)$.

Ballistic deposition typically yields a delicate forest
of closely packed trees.  The cluster is self-affine with surface
roughness that is believed to be in the KPZ universality class.

\subsection{Restricted solid-on-solid growth}
\label{su-sos}

The solid-on-solid model~\cite{MeRaSaBa,KiKo,KrSp} was originally
devised to study crystal growth.  The version presented here is most
closely related to the model studied in~\cite{KiKo} and referred to as
the `restricted solid-on-solid model' (RSOS).  We begin with an
undirected graph, $G = (V,E)$, and a maximum allowed height $H$\@.  The
RSOS clusters may be described by a set of column heights that can
only be incremented if they are less than or equal to the height of
all neighboring columns.  The initial cluster consists of a set of
sites together with their starting column heights (see below).  The
RSOS model can be viewed as a cluster growth model on a space-time
graph with directed bonds,
\begin{eqnarray}
\label{eq:rsosgraph}
E\s & = &
\{ ([i,h-1],[j,h]) \mid \{i,j\} \in E \; {\rm and} \; h=1, 2, \ldots, H \} \\
& & \;\;\;\;\; \cup \; \; \; \{ ([j,h-1],[j,h]) \mid j \in V \; {\rm and} \;
h=1, 2, \ldots, H \}.
\nonumber
\end{eqnarray}

\noindent
Technically, if the set of sites in the initial cluster $S$ consists of
$i$, $1 \leq i\leq n$, with corresponding initial heights $h(i,0)$,
then \[ S = \{ [i, x] \; | \; 1 \leq i \leq n \; {\rm and} \; 1 \leq x \leq
h(i,0) \}
\cup \{ [i,0] \; | \; \forall \; i \in V \} \] forms the initial
cluster in the space-time graph.  Here $h$ is the height function
described previously.  Typically, the first component of $S$ will be
empty.  Allowed growth sites on the space-time graph have the property
that {\em all\/} of their immediate predecessors are in the cluster.
At each step, an allowed growth site is randomly chosen and added to
the cluster.

Though the interfaces created by RSOS growth are locally smoother than
those created by ballistic deposition and Eden growth, it is believed
that the scaling of the interfacial width is in the KPZ universality
class.

\subsection{Waiting time growth models and minimum weight paths}
\label{su-kpz}

The close connection between random growth models and minimum weight
paths is exploited in the parallel algorithms discussed in
Section~\ref{se-parallelgrowth}.  To see this connection consider a
waiting time growth model~\cite{RoHaHi,TaKeWo} in which a random
waiting time, $\tau_j$, is independently assigned to each site, $j$,
of the graph.  Growth starts from an initial cluster $S$\@. The
cluster grows with a real-valued time parameter, $t$, and site $j$ is
added to the cluster at time $t_j$ according to the rule
\begin{equation}
\label{eq:wgrow}
t_j = \tau_j + \min_{\{k,j\}\in E} \; [ t_k].
\end{equation}
Each site $l$ in $S$ has $t_l = 0$.  The minimization gives the time
that $j$ is added to the perimeter.  The growth rule is that site $j$
waits a time, $\tau_j$, from the time it joins the perimeter to the
time it is admitted to the cluster.
By iterating Eq.~(\ref{eq:wgrow}), one sees that $t_j$ is in fact the
weight of a MWP from $S$ to $j$ (using site waiting times as weights),
\begin{equation}
\label{eq:wpath}
t_j=\min_{ \Gamma}[\,\sum_{k \in \Gamma} \tau_k\,].
\end{equation}
The minimization is over all simple paths, $\Gamma$, from an element
of $S$ to site $j$.  Note the weight of the source site is not
included in the path weight but the weight of the destination site is.
The mapping between growth models and MWPs is the discrete analog of
the transformation between the KPZ equation and the directed polymer
in a random environment~\cite{KPZ,KaZh}.

The waiting time model defines an ordering function, $c(n)$, given by
sorting the sites by the times they are added to the cluster.  The
statistics of this ordering depend on the probability law, $P(\tau)$,
for the waiting times. By an appropriate choice of the waiting time
distributions, one can recover Eden growth, solid-on-solid growth or
invasion percolation.  For the Eden model~\cite{RoHaHi} the waiting
times are chosen from the exponential distribution; for the RSOS
model~\cite{TaKeWo} the waiting times are chosen from a negative
exponential distribution,\[P(\tau)=\exp(\tau)\] with $\tau <0$.  In
both cases, the `no memory' property of the exponential distribution
ensures that the next site added to the cluster is chosen at random
from among all potential growth sites.

More generally, if the tail of $P(|\tau|)$ decays sufficiently fast
the cluster is compact with a self-affine surface and is believed to
be in the KPZ universality class.  Recent
studies~\cite{Zh,TaKeWo,Kr,RoHaSi} have shown that new universality
classes result from distributions with long tails.  We show here that
if the waiting time distribution is sufficiently broad and the growth
is on an undirected graph, the result is invasion percolation
clusters.

Invasion percolation clusters can be obtained from a waiting time
growth model with waiting times that are broadly distributed.  Let
$\tau_j$ equal $2^{n_j}$ with $n_j$ an integer-valued random variable
chosen uniformly on $1,2,\ldots,M$ (the value of $M$ that is necessary
is discussed below).  The ordering obtained from the waiting time
model will agree with the invasion percolation order described in
Section~\ref{su-ip} when site $j$ receives a weight of $x_j$ equal to
$n_j/M$. Suppose that all of the $n_j$'s are distinct so that the
cluster order is unambiguous. The sum of the waiting times,
$t_\Gamma$, along a given path, $\Gamma$, can be written in binary
with the $n_k{}^{\rm th}$ place a `$1$' if $k$ is on the path and a
`$0$' otherwise.  It is easy to see that two paths, $\Gamma$ and
$\Gamma'$, are equal if and only if $t_\Gamma$ equals $t_{\Gamma'}$.
Furthermore, if $|t_\Gamma - t_{\Gamma'}| < \tau_k$ then site $k$ is
on path $\Gamma$ if and only if it is on path $\Gamma'$.

The proof of the equivalence of the two models for invasion
percolation, assuming distinct $n_j$'s, is by induction.  Clearly the
first site added to the cluster is the same for both models. Suppose
that the invasion percolation order, $c(\cdot)$, agrees with the
waiting time order, $d(\cdot)$, for the first $m$ steps.  We must show
that the next site added to the invasion percolation cluster agrees
with the next site added to the waiting time cluster,
i.e. $c(m+1)=d(m+1)$.

Let $i=c(m+1)$ and $j=d(m+1)$.  Both $i$ and $j$ are on the perimeter
of the cluster at step $m$ and for any other site $k$ on the
perimeter, $\tau_i < \tau_k$ and $t_j < t_k$.  Let $i'$ denote the
first neighbor of $i$ that was added to the cluster, so that $t_{i'} =
t_i-\tau_i$. By definition of the waiting time model, it follows that
$t_j > t_{i'}$.  Combining with the previous statement, we can write
$t_i - t_j < \tau_i$.  Since $j$ is the next site added to the waiting
time cluster, we have $t_j \leq t_i$ so that $|t_i-t_j |<
\tau_i$. From this inequality it follows that $i$ is on the MWP to
$j$.  However, since $i$ is the next site added to the invasion
percolation cluster, we have $\tau_i \leq \tau_j$ leading to the
conclusion that $j$ is also on the MWP to $i$.  Thus, $i$ equals $j$
and the induction is complete.  For further discussion and proof
details see~\cite{GrMatechreport}.

Invasion percolation is not quite in the form of a waiting time growth
model because, for finite values of $M$, the probability of two sites
having the same value of $n_j$ is non-zero.  If $M$ is made
sufficiently large the probability of a tie can be made small but the
limit as $M$ approaches infinity is not a well-defined probability
distribution.  Thus the waiting time model with independent waiting
times does not always generate a well-defined cluster ordering.  On
the other hand, if the $n_j$ are chosen to be a random permutation of
$N$, the number of sites in the graph, then the invasion percolation
distribution is exactly reproduced.  To see this suppose that $x_j$ is
chosen as an uniform random deviate as described in
Section~\ref{su-ip}.  If the $x_j$'s are sorted in ascending order and
$n_j$ is the rank of $x_j$, then the distribution of the $n_j$'s is a
random permutation on $N$\@.  In this case the order induced by the
waiting time growth model with $\tau_j$ equal to $2^{n_j}$ is the same
as the invasion percolation order for the $x_j$'s.

\section{Fast parallel algorithms for growth models}
\label{se-parallelgrowth}

The probabilistic parallel algorithms presented below generate cluster
orderings (and/or height functions) for each growth model.  A single
run of any of the algorithms requires polylog time and uses a
polynomial number of processors on a probabilistic CREW P-RAM\@.  From
the analysis of the algorithms, we can conclude that there are natural
decision problems based on them that are in the complexity class {\bf
RNC}$^2$.

The input to the algorithms is an undirected graph, $G = (V,E)$, and
an initial cluster, $S$\@.  For Eden growth and invasion percolation,
$S$ is a connected subset of $V$\@.  For ballistic deposition and the
RSOS model, $S$ is a connected subset of the space-time graph that is
an allowed cluster according to the rule of the given model.  In each
case, let $N$ equal $|V|$.  The algorithms use one of the following
two strategies and are grouped appropriately.

\begin{description}

\item[Random Weight]
The first group relies on the equivalent waiting time model discussed
in Section~\ref{su-kpz}.  The randomization step assigns random
weights to the sites.  MWP problems are then solved to simulate the
growth models in parallel.  This approach is applied to Eden growth,
the RSOS model and invasion percolation with and without trapping.

\item[Random List]
The second group relies on a strategy in which a random list of sites
is generated. A site in the list is added to the cluster at a given
step if it is an allowed growth site of the cluster generated by
the previous elements of the list.  The MWP algorithm is used to
determine in parallel whether elements of the list are allowed growth
sites.  Algorithms using this approach are given for Eden growth and
the RSOS model.  A ballistic deposition algorithm is also included
under this paradigm.

\end{description}

We present two separate algorithms for Eden growth and the RSOS model
because of differences in efficiency and precision. For these two
models the random weight approach is faster, and uses processors and
random numbers more efficiently but it {\em approximates\/} the correct
cluster distribution.  The random list approach generates the exact
cluster distribution.  Note that with small probability either type of
algorithm may fail to produce the desired output.  Thus it may be
necessary to make repeated trials.

\subsection{Algorithms based on random site weights}
\label{su-prsw}

The structure of the parallel algorithms in this group may be summarized as
follows:

\begin{tabbing}
{\bf Random Weight Paradigm}\\
\hspace*{.25in} \= {\bf beg}\={\bf in}\\

\>\>1. In parallel, assign each site $j$ a random weight, $w_j$.\\

\>\>2. In parallel, for each site $j$ assign $W_j$ the weight of the MWP
from $S$ to $j$.\\

\>\> 3.
In \= parallel, sort the sites by $W_j$ in increasing order. Assign
$c(n)$ the $n^{\rm th}$ element in the list \\
\>\>\> of sorted sites.  If two
sites have the same path weight, the algorithm fails.\\

\hspace*{.25in} {\bf end}.
\end{tabbing}

We analyze the time and processor complexity of the algorithms on a
CREW P-RAM\@.  Assuming there are at least $N$ processors, step one
requires constant time.  The second step requires $O(\log^2 N)$ time
and $N^3 /\log N$ processors (see Section~\ref{su-mwpath}).  The third
step can be performed in $O(\log N)$ time using $N$
processors~\cite{GiRy}.  Thus, the overall resource bounds for the
procedure are $O(\log^2 N)$ time and $N^3 / \log N$ processors.  We
will have more to say about each growth model as it is considered in
turn.  The specific choice of random weights for each model is given
below.

\subsubsection{Eden growth (I)}
\label{su-eden1}

The waiting time growth model yields Eden growth if the waiting times
are chosen from an exponential distribution.  If $n_j$ is chosen
uniformly on $1, 2,\ldots,M$ and $w_j$ is assigned $\log M-\log n_j$
then $w_j$ approximates a random deviate chosen from the exponential
distribution.  The path weight is the sum of the site weights
(excluding the source but including the destination) along the path
and all of the numerical operations are done with $O(\log M)$-bit
numbers.  It suffices to choose $M=\Omega(N^3)$ to make the failure
probability much less than one, thus $O(N\log N)$ random bits are
required.\footnote{$f(n)$ is $\Omega(g(n))$ if there exists a constant
$c > 0$ and a natural number $N_0$ such that for all $n > N_0$,
$|f(n)| \geq c|g(n)|$.}  This is the same order as the defining
sequential algorithm, see Section~\ref{su-eg}.

Since the algorithm has a non-vanishing failure probability, it may
have to be run several times to generate a single cluster.  The time
for one run of the algorithm is $O(\log^2 N)$.  The time to generate a
complete cluster is not bounded.  Since waiting times that are
narrowly distributed lead to more compact clusters and are more likely
to fail with finite precision arithmetic, the algorithm presented here
does not sample the exact distribution Eden cluster except in the
limit as $M$ approaches infinity.

\subsubsection{Restricted solid-on-solid (I)}
\label{su-rsos1}

For the RSOS model growth occurs on the directed space-time graph,
$G\s$, defined in Section~\ref{se-growth} and in
Eq.~(\ref{eq:rsosgraph}).  Here the waiting times are chosen from a
negative exponential distribution.  This is implemented by choosing
$n_j$ uniformly on $1,2,\ldots,M$ as before and letting $w_j$ be $\log
n_j - \log M$.  The minimization over paths is restricted to {\em
directed\/} (simple) paths from $S$ to $j$.  The running time of the
algorithm is the same as that for Eden growth.  The algorithm
approximates the RSOS distribution for finite $M$.

\subsubsection{Invasion percolation}
\label{su-pip}

As discussed in Section~\ref{su-kpz}, invasion percolation is
equivalent to a waiting time model in which the logarithms of the
waiting times are a random ordering of the sites.  That is, $w_j$ is
$2^{\Pi[j]}$, where $\Pi$ is a random permutation of $1, 2, \ldots,
N$\@.  Fast parallel algorithms for generating random permutations are
given in~\cite{Ha} and are discussed in Section~\ref{su-ppram}.
Unlike Eden growth and the RSOS model, the algorithm samples the exact
invasion percolation distribution in $O(\log^2 N)$ time since no two
distinct paths can have the same path weight.\footnote{Note, the time
bound given for invasion percolation is for the {\em uniform cost\/}
model that is used throughout this paper.  The numbers involved in the
algorithm are actually very large although they are still only a
polynomial number of bits in the input size.  Since the sum and
minimum of two $N$-bit numbers can be performed in {\bf NC}$^1$, if we
analyzed the algorithm in a straightforward way using the {\em log
cost\/} model an additional log factor would appear in the running
time.  An alternative implementation is known that guarantees the
algorithm runs in $O(\log^2 N)$ even using the log cost model.}

\subsubsection{Invasion percolation with trapping}

The trapping rule does not affect the order in which sites are tested
for being in the cluster. The strategy for generating invasion
percolation with trapping is to first generate an invasion percolation
cluster without trapping and then to test each site for being trapped.
We describe an algorithm for performing this computation.

\newpage

\begin{tabbing}
{\bf Parallel Random Weight Invasion Percolation with Trapping Algorithm}\\
\hspace*{.25in} \= {\bf beg}\={\bf in}\\

\>\> 1. $L$ $\leftarrow$ $V$.\\

\>\> 2. Compute an invasion percolation order, $c(\cdot)$, for $G$.\\

\>\> 3. Fo\=r each site $i$, $inv(i) \leftarrow c^{-1}(i)$.\\
\>\>\>(Note that $inv(i)$ is the step at which site $i$ is added to the
cluster.)\\

\>\> 4. For each site $i$ in parallel $w_i$ $\leftarrow$ $2^{-inv(i)}$.\\

\>\> 5.
$W_i$ $\leftarrow$ the MWP from $i$ to the sink,
$s^{\prime}$, where the path is computed using the above site\\
\>\>\>weights and the weight of $i$ itself is not counted.\\

\>\> 6. If $W_i$ $>$ $2^{-inv(i)}$ then  delete $i$ from $L$.\\

\>\> 7.
Sort the remaining (untrapped) sites in $L$ based on their invasion
percolation labels,\\
\>\>\>$inv(i)$, and let $c(n)$ be the $n^{\rm th}$ element in the list.\\

\> {\bf end}.

\end{tabbing}

\noindent
For the sake of clarity, we have not treated the initial cluster as a
special case.  It is straightforward to take care of this.  For
example, when $i$ is 0, $c^{-1}(i)$ represents the initial cluster.
The complexity of this algorithm is the same as that of the one
described above for invasion percolation.  Additional details and the
proof of correctness for the algorithm can be found
in~\cite{GrMatechreport}.

\subsection{Algorithms based on random lists of sites}
\label{su-prl}

The algorithms for the Eden and RSOS models presented below
parallelize the following sequential strategy. At each step a site is
chosen at random.  If the site is an allowed growth site, it is added
to the cluster; otherwise, it is discarded.  To parallelize this
method, the randomly chosen sites are first prepared as a list and
then the MWP algorithm is used to determine whether each is an allowed
growth site.  Since most randomly chosen sites are not legal growth
sites, this method is inefficient in its use of randomness and memory.
However, unlike the algorithms of Sections~\ref{su-eden1}
and~\ref{su-rsos1}, these algorithms sample the exact cluster order
distribution for the Eden and RSOS models.  A variant of this approach
is used to sample the height function distribution for ballistic
deposition.

\subsubsection{Eden growth (II)}
\label{su-peg}

The pseudo-code given below produces Eden clusters in parallel. The
output of the algorithm is the cluster order. Explanatory remarks are
given after the algorithm.

\begin{tabbing}
{\bf Parallel Eden Growth (II) Algorithm}\\
\hspace*{.25in} \= {\bf beg}\={\bf in}\\

\>\>1. Gen\=erate a random list of sites, $(v_1,v_2,\ldots,v_T)$, chosen from
$V - S$.\\

\>\>2. Construct a directed graph, $G'=( \{0,1,\ldots,T\},E')$,
where\\

\>\>\> for each $1 \leq m,n \leq T$, if $(\{v_m,v_n\} \in E$ and $m<n)$ then
$(m,n) \in E'$, and\\

\>\>\> for each $1 \leq n \leq T$, if $v_n$ is on the perimeter of $S$ then
$(0,n) \in E'$.\\

\>\>3.
Let $r$ be the number of sites in $G'$ that can be reached from site 0
by a simple
path.\\

\>\>4.
Let $R[k] = j$, where $j$ is the $k^{\rm th}$ site in $G'$ (in sorted order)
reachable from site 0 \\

\>\>\> ($1 \leq k \leq r$ and $j \in \{ 1, 2, \ldots, T \}$).\\

\>\>5. For $k \leftarrow 1$ to $r$ do $L[k] \leftarrow v_{R[k]}$.\\

\>\>6. Compact $L$ by deleting all but the first
appearance of each site.\\

\>\> 7. $c(n) \leftarrow L[n]$ is the $n^{\rm th}$ site added to the cluster
($n$ starts at 1).\\

\> {\bf end}.

\end{tabbing}

The construction in Step~2 insures that $G'$ reflects both the
connectivity of the original graph $G$ and the order of adding sites
established by the list $(v_1,v_2,\ldots,v_T)$.  If an element $m$ is
connected to site 0 in $G'$ then its pre-image, $v_m$ is either a
perimeter site at step $m-1$, or it has already joined the cluster
before step $m$.  If $m$ is connected to site 0, its pre-image is
added to array $L$ in Step~5.  A site becomes part of the cluster when
it makes its first appearance in $L$\@. In Step~6 additional
appearances of every site are deleted from $L$ resulting in the
cluster order.  Since the original list was chosen at random, each
successive site in the cluster is chosen at random from the allowed
growth sites as required by the definition of the Eden model.

As presented the algorithm has a non-vanishing probability of failing
to produce a cluster of a given size.  The probability of obtaining a
complete cluster of size $K$ equal to $N$ can be made close to one
with the choice of $T=\Omega(N^2)$.  This is the case since there is
at least one perimeter site available at each time step and the
probability of choosing that site is at least $1/N$.  If failures are
permitted, the Eden cluster distribution is not perfectly sampled since
clusters with atypically large perimeters are less likely to fail and
will be favored.  However, one can generate the exact distribution by
iterating the algorithm several times using the final cluster from one
run as the initial cluster for the next run.  The iteration is
continued until a cluster of the desired size is obtained.  This
method produces an unbiased sample of Eden clusters.

For $d$-dimensional lattices Eden clusters are compact~\cite{Ri}, so
it is possible to estimate how large $T$ must be and thus how many
random numbers are required to make the failure probability small. The
perimeter of a $d$-dimensional cluster of size $N$ is expected to
scale as $O(N^{\frac{d-1}{d}})$, so that it suffices to choose
$T=\Omega(N^{2-\frac{d-1}{d}})$ to insure that the perimeter is hit
$N$ times.  Therefore, the algorithm needs $O(N^{1+\frac{1}{d}}\log N)$
random bits.  This contrasts with the requirement of $O(N\log N)$
random bits for the waiting time algorithm of Section~\ref{su-eden1}.

All of the steps in the algorithm are straightforward to parallelize
using techniques like parallel sorting and parallel prefix
computations~\cite{KaRa}.  Each requires a polynomial number of
processors.  For a choice of $T$ equal to $N^2$, the running time of
the algorithm is dominated by the parallel minimum path subroutine
used in step~3 (to test reachability) and is $O(\log^2 N)$.  As noted
above, with small probability the algorithm may not produce a cluster
of the desired size in a bounded number of runs.  The expected running
time to produce a complete cluster is $O(\log^2 N)$.

\subsubsection{Restricted solid-on-solid (II)}
\label{su-psos}

The algorithm given below simulates the RSOS model in parallel.  This
algorithm builds and then operates on the space-time graph.  The
initial cluster $S$ consists of all sites $[i,h]$ with $h \leq
h(i,0)$, where $h$ with two arguments is the height function.  We
assume the initial heights are $O(N)$ and have a maximum achievable
height of $H$\@.  In most cases the initial heights are taken as zero
so that $S=\{[i,0] \mid i \in V\}$.  The output of the algorithm is
the space-time cluster order.  The height function can easily be
obtained from the cluster order.

\newpage

\begin{tabbing}
{\bf Parallel Restricted Solid-on-solid (II) Algorithm}\\
\hspace*{.25in} \= {\bf beg}\={\bf in}\\

\>\>1. Bui\=ld a directed space-time graph $G\s=(V\s,E\s)$, where\\

\>\>\>$V\s \leftarrow \{ [i,h] \mid i \in V \;
{\rm and} \; h=0,1, \ldots, H\}$, and\\

\>\>\> $E\s \leftarrow \{ ([i,h-1],[j,h]) \mid \{i,j\} \in E$ and $h=1, 2,
\ldots, H \}$ \\

\>\>\> \hspace*{.6in} $\cup \; \; \{ ([j,h-1],[j,h]) \mid j \in V$ and $h=1, 2,
\ldots, H \}$.\\

\>\>2.  Generate a random list of size $T$ of elements from $V\s$,
$L[m] \leftarrow [v_m,h_m]$, ($1 \leq m \leq T$). \\

\>\>3. Construct a directed graph, $G'=( \{0,1,\ldots,T\},E')$,
where\\

\>\>\> for each $1 \leq m,n \leq T$, if $(([v_m,h_m],[v_n,h_n] )\in E\s$
and $m<n)$ then $(m,n) \in E'$, and\\

\>\>\> for each $1 \leq n \leq T$, if $[v_n,h_n]$ is on the perimeter of
$S$ then $(0,n) \in E'$.\\

\>\>4. For \= $k \leftarrow 1$ to $T$ do
$A'[k] \leftarrow \{n\mid \exists $\/ a directed path from $n$
to $k$ in $G'\}$.\\

\>\>\>(Note, $A'[k]$ represents the set of ancestors of $k$ in $G'$.)\\

\>\>5. For $k \leftarrow 1$ to $T$ do $B[k] \leftarrow \{[i,h]\mid \exists $\/
a directed path from
$[i,h]$ to $L[k]$ in $G\s\}$. \\

\>\>\>(Note, $B[k]$ represents the set of ancestors of $[v_k,h_k]$ in
$G\s$.)\\

\>\>6. For \= $k \leftarrow 1$ to $T$ do
$A[k] \leftarrow \{ [i,h] \mid  n \in A'[k]$ and $ [i,h] = L[n]\}$.\\

\>\>\> (Note, $L[0] = S$.)\\

\>\>7. For $k \leftarrow 1$ to $T$ do if $B[k] = A[k]$ then $L[k]\leftarrow
L[k]$ else $L[k]\leftarrow -1$.\\

\>\>8. Compact $L$ by deleting $-1$'s. \\

\>\>9. Compact $L$ by deleting all but the first
appearance of each site.\\

\>\> 10. $c(n) \leftarrow L[n]$ is the $n^{\rm th}$ site added to the
cluster ($n$ starts at 1).\\

\>{\bf end}.
\end{tabbing}

\noindent
Note that the initial cluster needs to be treated as a special case.

A random list of space-time sites, $L$, is generated in Step~2 and a
directed graph, $G'$, is constructed from this list.  A directed bond
exists in $G'$ if it corresponds to a directed bond in $G\s$ and if it
is compatible with the ordering of the list $L$\@. The elements of $L$
are potential growth sites and are accepted into the growing cluster
if all of their ancestors in $G\s$ are already in the cluster.
Acceptance in the cluster is permitted if the set of ancestors of a
site in $G'$ includes all the ancestors of the pre-image of that site
in $G\s$\@. All elements that fail this test are deleted from $L$\@.
The parallel MWP algorithm is used to determine ancestors by testing
connectivity.  The ordering of $L$ determines the cluster order.

Several remarks follow regarding the algorithm.  First, as is the case
for the parallel Eden growth~(II) algorithm, the parallel RSOS~(II)
algorithm may fail before the desired cluster size is obtained.  If
failures are allowed, smooth interfaces will be favored over rough
interfaces because smooth interfaces have more allowed growth sites.
Thus, the RSOS distribution will not be exactly sampled. If the exact
distribution is required, it will occasionally be necessary to iterate
the algorithm using the height function from one run as the initial
cluster for the next run.  Although this procedure is not guaranteed
to halt in a fixed time, the probability of failure can be made small
so that the expected running time of the iterated algorithm is
polylog.

Second, the algorithm may need $O(N\s^2 \log N\s)$ random bits to form
a complete cluster, where $N\s=N(H+1)$ is the number of sites in the
space-time graph.  This is because each element in the list has a
chance of at least $1/N\s$ to be a growth site and growth must occur
$N\s$ times.  This is more than the requirement of $O(N\s \log N\s)$
random bits for RSOS~(I), see Section~\ref{su-rsos1}.

Third, all of the steps in the algorithm are straightforward to
parallelize using techniques like parallel sorting and parallel prefix
computations~\cite{KaRa}.  Each requires a polynomial number of
processors in $N$\@.  For a choice of $T$ equal to $ N^2$, the running
time of the algorithm is dominated by the parallel MWP subroutine used
in Steps~4 and~5 (to determine ancestors) and is $O(\log^2 N)$.  As
noted above, with small probability the algorithm may not produce a
cluster of the desired size.

\subsubsection{Ballistic deposition}
\label{su-pbd}

The algorithm described below simulates ballistic deposition in
parallel.  The output of the algorithm is a realization of the height
function, $h(i,n)$, at site $i$ at time step $n$, $n=0, 1, \ldots, K$,
as given by the defining algorithm for ballistic deposition.

\begin{tabbing}
{\bf Parallel Ballistic Deposition Algorithm}\\
\hspace*{.25in} \= {\bf beg}\={\bf in}\\

\>\>1. Gen\=erate a random list of sites $(v_1, v_2, \ldots,v_K)$.\\

\>\> 2.  In parallel create a directed space-time graph $G\s=(V\s,E\s)$,
where\\

\>\>\>$V\s \leftarrow \{ [i,n] \mid i \in V \; {\rm and} \;
n=0,1, \ldots, K\}$ and\\

\>\>\> $E\s \leftarrow \{ ([i,n-1],[j,n]) \mid \{i,j\} \in E$ and $n=1, 2,
\ldots, K \}$ \\

\>\>\>\hspace*{.1in} $\cup \; \; \{ ([j,n-1],[j,n])
\mid j \in V$ and $n=1, 2, \ldots, K \}$. \\

\>\>3. Assign weights to the edges of the space-time graph as follows:\\

\>\>\>for \= each $j \in V$ and $1 \le n \le K$ do\\

\>\>\>\>if $j=v_n$ then $w(([j,n-1],[j,n])) \leftarrow 0$
else $w(([j,n-1],[j,n])) \leftarrow 1$, and\\

\>\>\>for each $\{i,j\} \in E$ and $1 \le n \le K$ do\\

\>\>\>\>if $j=v_n$ then $w(([i,n-1],[j,n])) \leftarrow 1$ else
$w(([i,n-1],[j,n])) \leftarrow \infty$.\\

\>\>4. for \= each $j \in V$ and $1 \le n \le K$ do\\

\>\>\>\>$h(j,n) \leftarrow$ $n$ $-$ the weight of the MWP from a site in $\{
[i,0] \mid i \in V \}$ to $[j,n]$.\\

\>{\bf end}.
\end{tabbing}

As was the case for the RSOS~(I) algorithm, this algorithm builds and
operates on a space-time graph, however, growth does not occur on this
graph.  A random list of sites is generated in the substrate.  This
list determines the order columns are chosen for growth.  The MWP
subroutine can be used to compute the height on each column.

We argue the correctness of the algorithm below.  The proof is by
induction.  After one step of growth, only site $v_1$ should have
height one.  All other sites should have height zero.  Notice in
Step~4 of the algorithm, the MWP from $[v_1,0]$ to $[v_1,1]$ will be
zero because of the weighting in $G\s$.  Therefore, $h(v_1,1) = 1 - 0
= 1$ as required.  For sites other than $v_1$, the MWP will have
weight one and the computation in Step~4 correctly assigns them a
height of zero.

Assume for the induction hypothesis that after $l$ steps of growth, the
heights of all sites are correctly computed by the algorithm.
Consider step $l + 1$.  Suppose the MWP from $\{ [i,0] \; | \; i \in V
\}$ to $[v_{l+1}, l+1]$ contains $[v_{l+1},l]$.  Then the MWP to any
``neighbor'' of $[v_{l+1}, l+1]$ from $\{[i,0] \; | \; i \in V \}$ can
be at most one less than the MWP to $[v_{l+1},l+1]$.  In $G$ this
means a neighbor of $v_{l+1}$ can have height at most one greater than
$v_{l+1}$.  By the induction hypothesis the heights computed after $l$
steps were correct.  Step~4 has the effect of incrementing the height
of $v_{l+1}$ by one.  This correctly simulates ballistic deposition.

Suppose $[v_{l+1},l]$ is not on the MWP to $[v_{l+1},l+1]$.  This
implies there is a ``neighbor,'' $[v, l]$, of $[v_{l+1}, l+1]$ with
shortest MWP among neighbors such that the MWP to $[v,l]$ plus one is
the MWP to $[v_{l+1},l+1]$.  By the induction hypothesis
$h(v,l)$ was correctly computed.  According to the rules for ballistic
deposition, $h(v_{l+1},l+1)$ should be equal to $h(v,l)$.  The
arithmetic in Step~4 yields $(l$ $+$ $1)$ $-$(MWP to $[v,l]$ $+$ 1).
This is the same as $l$ $-$ MWP to $[v,l]$.  Note, this value is the
maximum height of a neighbor of $v_{l+1}$ in $G$ after $l$ growth
steps.  Thus the algorithm is correct.

In contrast to the growth rules for the Eden model and the RSOS model,
in ballistic deposition every site is an allowed growth site.  Thus a
random list of length $K$ yields a height function that is increased
for a single site at each step.  There are no failures and the maximum
possible resulting height is $K$\@.  The running time of the algorithm
is dominated by the parallel MWP subroutine.  If $K=O(N)$, the running
time is $O(\log^2 N)$.  Only a polynomial number of instances of the
MWP problem need to be solved in parallel in Step~4.  These are solved
on graphs larger than the original but still polynomial in size.
Therefore, the number of processors required by the algorithm is
polynomial.  Note, the randomness in this algorithm is used only in
generating the initial random list.

\section{Discussion}
\label{se-discussion}

We have presented fast parallel algorithms for the following growth
models: invasion percolation, Eden growth, ballistic deposition and
restricted solid-on-solid growth.  These algorithms exploit the close
connection between growth models and minimum weight self-avoiding
paths in random environments.  This connection also highlights the
similarities among all of the models.  All of the algorithms run in
polylog time using a polynomial number of processors, although several
of them have a small probability of failure.

While fast algorithms exist for the growth models discussed here,
there is no known fast parallel algorithm for sampling DLA clusters.
It has been shown~\cite{Mac93a} that a natural sequential approach for
DLA defines a {\bf P}-complete problem; there is almost certainly no
fast parallel algorithm based on this approach. On the other hand,
there could be alternative ways of simulating DLA and the {\bf
P}-completeness proof for one problem that generates DLA clusters on
random inputs does not rule out an average case polylog time algorithm
even if {\bf P}$\neq${\bf NC}\@.  Indeed, although we have
demonstrated here that Eden clusters can be generated in polylog time
there is alternative natural problem yielding Eden clusters on random
inputs that is {\bf P}-complete~\cite{MacUn}.  Despite these caveats,
we conjecture that no fast parallel algorithm exists for DLA; this
model seems fundamentally more complex than the models we consider.

The existence of parallel algorithms for simulating these models
establishes a polylog upper bound on the parallel time required to
simulate these models.  It seems plausible that models generating
statistically self-similar or self-affine clusters cannot be simulated
in less than logarithmic time.  Thus they are strictly more complex
than models that can be simulated in constant parallel time.  However,
just as is the case for DLA, it is a much more difficult question to
establish lower bounds on computational resources.  The general
question of establishing lower bounds on the resources needed for
sampling distributions is an interesting and difficult one.

\subsection*{Acknowledgements} Ray Greenlaw would like to thank John Bates
for several helpful discussions.

\end{document}